# The Effect of Dopants on the Magnetoresistance of WTe$_2$

Steven Flynn, Mazhar Ali, and R. J. Cava*

*Department of Chemistry, Princeton University, Princeton New Jersey 08544, USA*

*Corresponding author: rcava@princeton.edu

**Abstract**

Elucidating the nature of the large, non-saturating magnetoresistance in WTe$_2$ is a significant step in functionalizing this phenomenon for applications. Here, Mo, Re, and Ta doped WTe$_2$ are compared to determine whether isovalent and aliovalent substitutions have different effects on the large magnetoresistance. By 1% substitution, isoelectronic doping reduces the magnetoresistance by a factor of 1.2 with an apparent linear trend, whereas aliovalent doping reduces the effect by over an order of magnitude while following a higher-order decay. The apparent increased sensitivity of the magnetoresistive effect to aliovalent doping over simple isoelectronic disorder supports the conclusion that the large magnetoresistance in WTe$_2$ arises from interactions between balanced hole and electron populations.

1. **Introduction**

Magnetoresistance (MR), the change in resistivity of a material due to an external magnetic field, is a well-studied and technologically useful property. Since its discovery, MR has been a topic of sustained interest in materials science studies and has been successfully incorporated into a number of applications including magnetic memory devices, magnetic sensors, and hard drives.[1-3] The MR phenomenon is typically divided into various categories, each characterized by the specific mechanisms that cause it or by the materials in which it is expressed. Ordinary MR generally refers to small changes in resistivity, on the order of a few per cent, observed in many non-magnetic elements.[4] Giant MR (GMR) and Colossal MR (CMR) refer to much larger effects typically observed in coupled, thin layers of ferromagnetic material and manganese-based perovskite materials respectively.[5-7] Extremely large, positive MR (XMR), recently observed in $WTe_2$ and several other non-magnetic compounds, is the subject of this study.[8-13] The magnitude of the XMR effect in $WTe_2$ and the conditions where it is observed (13,000,000% at 0.53 K and 60 Tesla applied field) in addition to the character of $WTe_2$ as a layered dichalcogenide, suggest that this material may improve upon current magnetic sensing technologies for some applications.[8]

A multitude of studies has recently been performed to elucidate the origin of XMR in $WTe_2$ and other materials[14-16], all supporting the presence of nearly equal hole and electron populations at low temperatures as the origin of the effect. However, an investigation of the effects of chemical manipulation of this n-p balance has yet to be reported. Here we investigate the effects of aliovalent and isoelectronic doping on the magnetoresistance in polycrystalline $WTe_2$. Tantalum and Rhenium are selected as aliovalent dopants because they are adjacent to W in the periodic table and are expected to provide nominally p-type (Ta doping) or n-type (Re doping) variations. We find that the magnitude of the XMR effect is much more strongly dependent on the doping concentration in both the aliovalent cases than in the isoelectronic case. Moreover, the apparent difference in trends of XMR vs. dopant concentrations for either *n* or *p* doping compared to Mo suggests that XMR suppression may occur through different mechanisms in the two cases.

## 2. Experimental

Polycrystalline samples of the form $W_{1-x}M_xTe_2$ (M = Mo, Re, Ta; $x$ = 0.00033, 0.001, 0.0033, 0.01, 0.02, 0.03) were synthesized from their elemental components using sealed tube synthesis. Stoichiometric amounts of powdered Tungsten (99.999%), purified Tellurium (99.99%), and either powdered Tantalum (99.98%), Rhenium (99.999%), or Molybdenum (99.9%) were combined in a quartz tube which was then evacuated and sealed. Up to 5% extra Te was included to account for its vapor pressure at reaction temperatures. Total sample mass ranged from 0.5 to 10 grams; samples with smaller doping contents were produced in greater quantities in order to reduce error in the amount of dopant measured. Sealed mixtures were reacted at 700 °C for 2- 4 days. Longer reaction times were employed for samples with greater mass. The resulting material was homogenized through grinding, pressed into a dense pellet, and returned to the furnace in a newly sealed and evacuated tube and reacted at 750 °C for 2-4 days. A third heating at 750 °C was employed after additional grinding and pressing in order to promote sample homogeneity. The purity of the samples was confirmed through powder x-ray diffraction patterns collected at room temperature using Cu *K*α radiation on a Bruker D8 Focus diffractometer with graphite monochromator.

Temperature-dependent resistivity measurements were taken from 300 – 2 Kelvin (K) on dense, sintered polycrystalline pellets with a Quantum Design Physical Property Measurement System (PPMS) using the standard four-point probe technique. The magnetic field-dependence of the resistivity was measured at 2 K in applied fields ($\mu_0 H$) of 0 – 9 Tesla (T) using the same technique. The magnetoresistance of the $W_{1-x}M_x Te_2$ samples was calculated according to the formula:

$$MR = (\rho(H) - \rho(0))/\rho(0)$$

where $\rho(H)$ is the measured resistivity in an applied magnetic field $H$ at 2K.

## 3. Results and Discussion

The absolute temperature-dependent resistivities for the pure $WTe_2$ and 1% Mo, Re, and Ta doped polycrystalline samples are shown in Figure 1. While all four samples exhibit decreasing resistivity with temperature, their behaviors differ significantly. In the undoped sample, the resistivity decreases from 0.88 mΩ cm at room temperature to 0.045 mΩ cm at 2 K, yielding a residual resistivity ratio (RRR) of 19.5. The Mo doped sample strongly resembles the pristine compound, decreasing from 0.88 to 0.043 mΩ cm, presenting a slightly larger RRR of 20.2, within error of the same behavior for polycrystalline pellets. Both the Ta and Re doped samples, however, show a significantly smaller decrease in resistivity over the measured temperature range: from 0.90 mΩ cm at room temperature to 0.15 mΩ cm at 2 K for Ta, and from 0.89 mΩ cm to 0.26 mΩ cm for Re. The RRR of these samples are 6 and 3.4 respectively, indicating significantly more carrier scattering as a result of the inclusion of the charged impurities Ta and Re.

The resistivities as a function of $\mu_0H$ from 0 to 9 T at 2 K are shown in Figure 2 for representative samples of all three dopants at different levels of doping. In general, the curves flatten with increasing dopant concentration, primarily due to simultaneous increases in the 0 field resistivity and decreases in the 9 T resistivity. The extent of flattening varies significantly between dopants. Figures 2a and b demonstrate that relative to pure $WTe_2$, the XMR is significantly decreased by 1% substitution with either Ta or Re. In contrast, the curve of the 1% Mo-doped sample appears to differ only slightly from that of the pristine material.

The 2 K XMR values for all samples were extracted from the field-dependent resistivity data and are displayed in Figure 3. The data present a clear distinction between the dopants employed: while the XMR vs. $x$ curves for the Ta and Re doped materials resemble each other in both magnitude and apparent shape, the XMR values for the Mo doped materials exhibit consistently higher values and a more linear trend with $x$. Over this range of compositions, the Mo samples exhibit a decrease in XMR from ~1100% in the pristine material to ~450%, while at 0.03% substitution, the XMR is reduced from 1100% to about 600% with Ta and 450% with Re. Thus the XMR is more sensitive to these dopants by several orders of magnitude.

The similarity of the rapid suppression of MR caused by Ta and Re doping and the dramatic difference of that behavior from the gradual decrease in the Mo-doping case (Figure 3) may reasonably be ascribed to one or two possible underlying causes. The first would be that all three dopants alter the XMR through the same mechanism, namely impurity scattering, which is simply enhanced in Re and Ta relative to Mo due to their aliovalent character. The second would be that the two observed trends occur primarily through independent mechanisms. Since Mo is isoelectronic to W, substitution of the former for the latter in $WTe_2$ should primarily increase impurity scattering, thereby lowering mobility and suppressing the XMR. The RRR data do suggest that isoelectronic Mo introduces less disorder into the system than the aliovalent dopants do. The data in Figure 3, however, which clearly show dramatic differences in the sensitivity and *x* dependence of the XMR for Ta or Re doping vs. Mo doping, suggest that it is the change in electron count for the aliovalent doping that has the much larger effect. A more consistent picture is formed if disturbing the n-p balance in WTe$_2$ is considered the primary mode of XMR suppression and mobility lowering is taken as secondary. As such, our results provide experimental support for the role of a near perfect n-p balance as the origin of the XMR in WTe$_2$.

## 4. Conclusion

Resistivity measurements on systematically doped WTe$_2$ have revealed a distinct difference between the linear decrease of MR by isovalent doping with Mo, and the more rapid, higher-order suppression caused by both of the aliovalent dopants, Ta and Re. This behavior is well-explained by a strong dependence on the balance between charge carriers within the material. As such, our results provide experimental support for the role of a near perfect n-p balance as the origin of the XMR in WTe$_2$.

**Acknowledgments**

Funding for this work was provided by Princeton University and the NSF MRSEC program, grant DMR-1420541.

Figures Captions:

**Figure 1 (color online)**: Temperature-dependent resistivity for undoped $WTe_2$ and $WTe_2$ doped with 1% Mo, Re, or Ta.

**Figure 2 (color online)**: Field-dependent resistivity measurements at 2 K for representative samples of a) Ta-doped $WTe_2$ b) Re-doped $WTe_2$ and c) Mo-doped $WTe_2$. Data for the undoped material is reproduced in each panel for clarity. Colors correspond to extent of doping: dark blue is x = 0, light blue is x = 0.001, green is x = 0.01, and red is x = 0.03

**Figure 3 (color online)**: Magnetoresistance of doped polycrystalline $WTe_2$ samples as a function of doping content. The lines are guides for the eye.

Figures:

Figure 1:

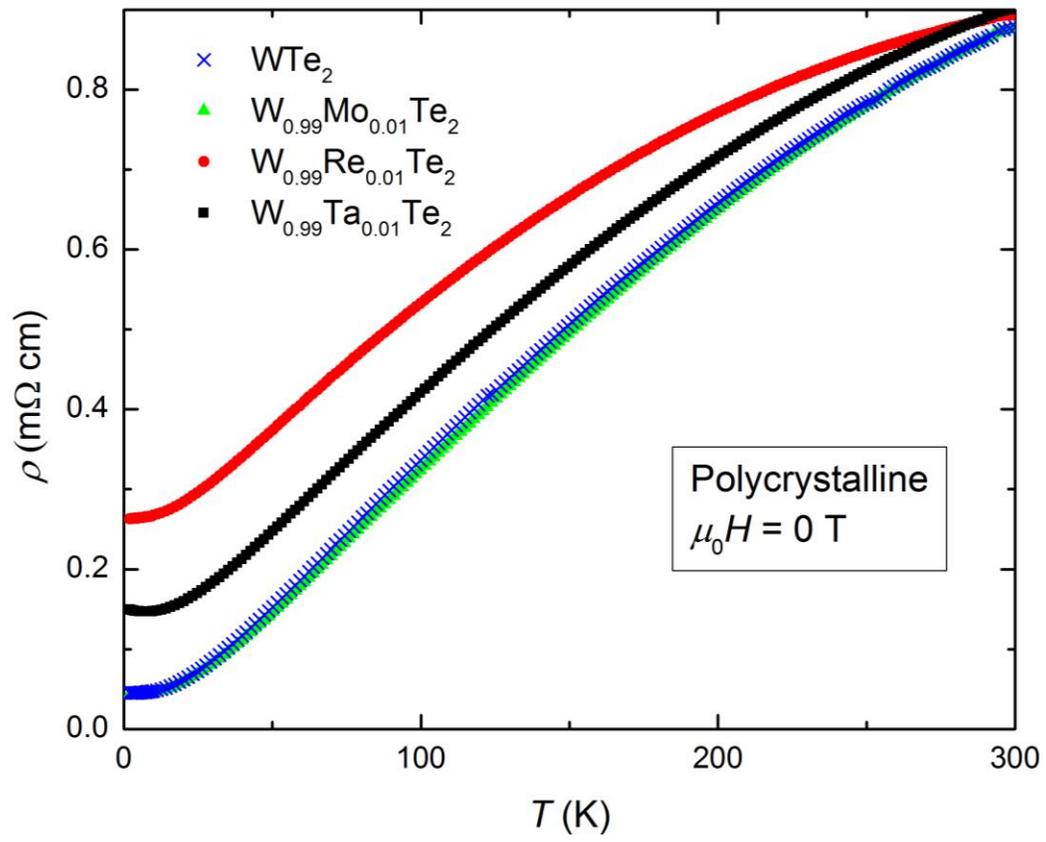

Figure 2:

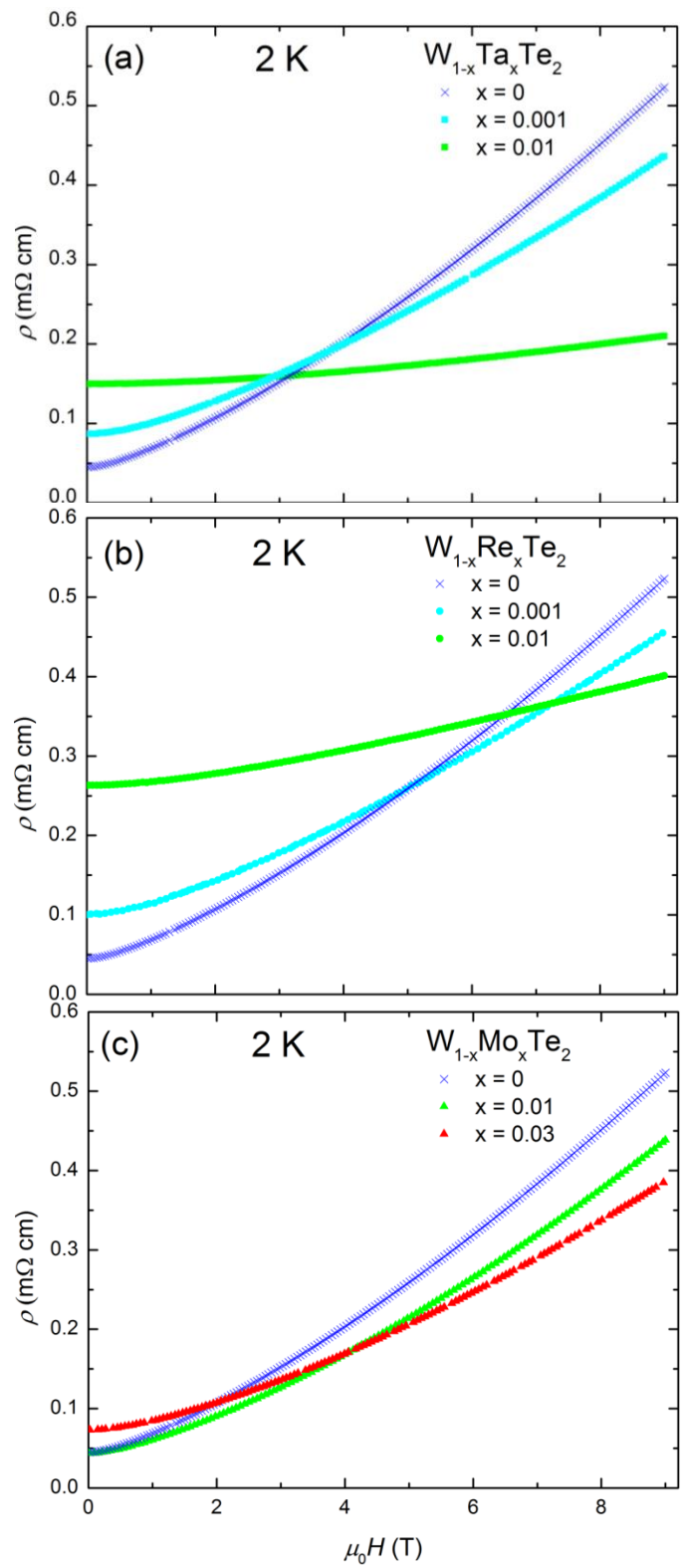

Figure 3:

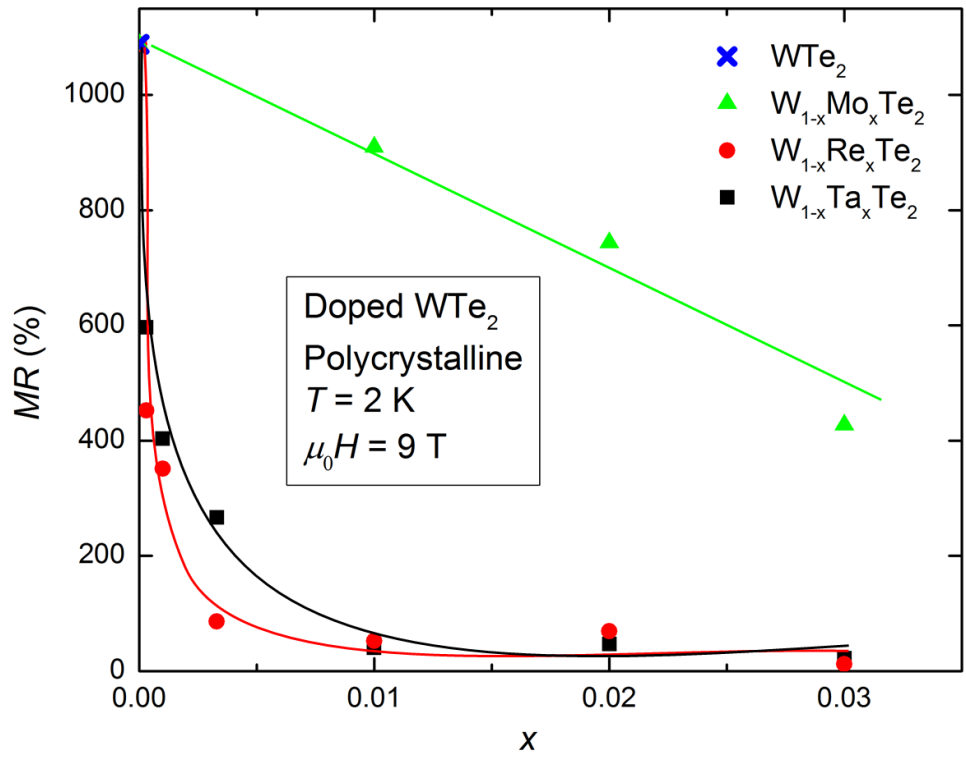